\documentclass[aps,prl,twocolumn,showpacs,preprintnumbers,amsmath,amssymb]{revtex4}

\usepackage{graphicx}% Include figure files
\usepackage{dcolumn}% Align table columns on decimal point
\usepackage{bm}% bold math

\allowdisplaybreaks

\newcommand {\beq} {\begin{equation}}
\newcommand {\eeq} {\end{equation}}
\newcommand {\beqa}{\begin{eqnarray}}
\newcommand {\eeqa}{\end{eqnarray}}

\newcommand{\idmat}{\mbox{\boldmath $1$}}

\begin{document}

%%%%%%%%%%%%%%%%%%%%%%%%%%%%%%%%%%%%%%%%%%%%%%%%%%%%%%%%%%%%%%%%%%%%
%  TITLE / AUTHOR                                                  %
%%%%%%%%%%%%%%%%%%%%%%%%%%%%%%%%%%%%%%%%%%%%%%%%%%%%%%%%%%%%%%%%%%%%

\title{Large $N$ reduction for Chern-Simons theory on $S^3$}

\author{Goro Ishiki$^{1}$}
\email{ishiki@post.kek.jp}
\author{Shinji Shimasaki$^{2}$}
\email{shinji@gauge.scphys.kyoto-u.ac.jp}
\author{Asato Tsuchiya$^{3}$}
\email{satsuch@ipc.shizuoka.ac.jp}

\affiliation{
$^{1}$
High Energy Accelerator Research Organization (KEK), 
Tsukuba, Ibaraki 305-0801, Japan \\
$^{2}$Department of Physics, Kyoto University, Kyoto 606-8502, Japan\\
$^{3}$Department of Physics, Shizuoka University,
836 Ohya, Suruga-ku, Shizuoka 422-8529, Japan
}

\date{August, 2009; preprint: KEK-TH-1326, KUNS-2226
%, hep-th/yymmnnn
%\today %%new
}% It is always \today, today,
             %  but any date may be explicitly specified

\begin{abstract}
We study a matrix model which is obtained by dimensional reduction of 
Chern-Simon theory on $S^3$ to zero dimension.
We find that expanded around a particular background consisting of 
multiple fuzzy spheres,
it reproduces the original theory on $S^3$ in the planar limit. 
This is viewed as a new type of 
the large $N$ reduction generalized to curved space. 
\end{abstract}

\pacs{11.25.Tq; 11.15.Tk; 11.10.Kk}

%11.25.Tq Gauge/string duality
%11.15.Tk Other nonperturbative techniques
%11.25.-w Theory of fundamental strings
%11.10.Kk Field theories in dimensions other than four  

%\pacs{11.25.-w; 11.25.Sq}
%11.25.-w Theory of fundamental strings
%11.25.Sq Nonperturbative techniques; string field theory

\maketitle

\paragraph*{Introduction.---}
Emergent space(-time) is often seen in recent developments
in string theory.
In particular, it is characteristic of
the matrix models \cite{matrix models} 
which are dimensionally reduced models of 
%obtained by dimensionally reducing 
ten-dimensional super Yang-Mills theory (SYM) 
%to lower dimensions 
and proposed as a non-perturbative formulation of 
superstring or M-theory.  While it is verified 
that these models reproduce gravity around flat space-time,
it is 
%not obvious
necessary to elucidate how curved space-time is included in these 
models \cite{curved space-time}.
%It can be said that 
The large $N$ reduction \cite{reduced model} developed decades ago
is the first example that realizes emergent space-time in matrix model. 
Indeed, it asserts that
the planar limit of gauge theory on flat space-time 
%in the planar limit 
is equivalent 
to the planar limit of a matrix model (a reduced model) that is
obtained by dimensional reduction to lower dimensions.
It is, therefore, worthwhile to study generalization of the large $N$ 
reduction to curved space-time. The large $N$ reduction can also give 
a non-perturbative formulation 
of planar gauge theory as an alternative to lattice gauge theory. 
%In this formulation, the matrix size plays a role of the ultraviolet cutoff.
It is well-known that in order to overcome the problem of flat directions (or the $U(1)^D$ symmetry breaking)
in the large $N$ reduction,
one needs the prescription \cite{Bhanot:1982sh}, 
%at least for the gauge theory on flat space-time in order to overcome
which is unfortunately not compatible with supersymmetry in gauge theory.
%so that the large $N$ reduction seems difficult to perform for supersymmetric gauge theory on flat space-time.

The authors of Ref. \cite{Ishii:2008ib} proposed a non-perturbative formulation of 
planar ${\cal N}=4$ 
%super  Yang-Mills theory (SYM) 
SYM on $R\times S^3$ equivalent 
to that on $R^4$ at a conformal point,
using the plane wave matrix model (PWMM) \cite{Berenstein:2002jq} 
%(For earlier discussions, 
(See also
\cite{Ishiki:2006yr,Ishii:2008tm}). PWMM is obtained by dimensional reduction of 
${\cal N}=4$ SYM on $R\times S^3$ to $R$ \cite{Kim:2003rza}, and in the formulation 
the $S^3$ is realized as a non-trivial fiber bundle
over $S^2$ by expanding PWMM around a particular background which 
%preserves sixteen supersymmetries and 
consists of multiple fuzzy spheres.
Thus the formulation is viewed as a new type of the large $N$ reduction generalized to curved space. 
The formulation overcomes the aforementioned problem in 
the large $N$ reduction for supersymmetric gauge theory
%because 
%there are no flat directions due to 
thanks to massiveness
%the curvature of the original $S^3$ 
and supersymmetry of PWMM \cite{lattice susy}.
By putting the formulation on a computer in terms of the method \cite{Anagnostopoulos:2007fw}, 
%for instance, 
one should be able to
study the strongly coupled regime of ${\cal N}=4$ SYM and therefore to
perform new non-trivial tests  for
%not done so far for
%for the sectors not protected by supersymmetry in
the 
%conjectured 
AdS/CFT duality \cite{Maldacena:1997re}.

A reasonable argument 
%strongly 
supporting the validity of the proposal was given in \cite{Ishii:2008ib}, and
the proposal has already passed 
some non-trivial tests at weak coupling \cite{Ishii:2008ib,Ishiki:2008te}.
%\cite{application}.
Application of the same type of the large $N$ reduction
to various gauge theories was also discussed in \cite{Hanada:2009kz}. 
However, the proposal has not been proved completely yet, in particular at strong coupling. 
In this letter, we prove that the same type of the large $N$ reduction does work even at strong coupling
for another gauge theory, Chern-Simons (CS) theory on $S^3$, 
which has been exactly 
solved \cite{Witten:1988hf}.
Our study is an extension of the 
%earlier 
classical analysis done 
in \cite{Ishii:2007sy} to the quantum level and
provides the first proved example of the new type of the large $N$ reduction.
% CS theory on $S^3$ is an interesting topological field theory 
% associated with the knot theory
%and therefore interesting in its own right.
Our formalism gives a new quantum mechanical definition of
%regularization method for 
%planar 
CS theory on $S^3$, which is an interesting 
topological field theory associated with the knot theory.
CS theory on $S^3$ is also interpreted as open topological A strings 
on $T^{\ast}S^3$. This aspect 
%is also expected to 
should enable us to gain from our findings 
%some 
insights into
%relate our findings to
%in the presence of $N$ D-branes wrapping $S^3$. 
%Topological strings 
%capture some aspects of more realistic string. 
%Thus lso in that sense
%expected to give some insights into 
the study of formulating superstring
non-perturbatively in terms of matrix model. 
%as non-perturbative formulation of superstring or M-theory.

\paragraph*{CS theory on $S^3$.---}
Let us review some exact results for $U(N)$ CS theory on $S^3$,
whose action is given by
%The action of the theory is given by
\begin{align}
S_{CS}=\frac{k}{4\pi}\int_{S^3}\mbox{Tr}\left(A\wedge dA+\frac{2}{3}A\wedge A\wedge A\right).
\label{Chern-Simons theory on S^3}
\end{align}
%where $k$ must be an integer.
The partition function of the theory, ${\cal Z}_{CS}$,
% \begin{align}
% {\cal Z}_{CS}=\int {\cal D}A\; e^{iS_{CS}}
% \label{partition function of CS}
% \end{align}
defines a topological invariant of the manifold $S^3$.
%which also depends on a choice of framing.
Given an oriented knot ${\cal K}$ in $S^3$, one can consider the Wilson loop in an irreducible representation
$R$ of $U(N)$
\begin{align}
W_R({\cal K})
=\mbox{Tr}_R \left(P\exp \oint_{{\cal K}}A \right). 
%=\mbox{Tr}_R \left(P\exp \int_0^1 A_M(z(\sigma))\frac{dz^M(\sigma)}{d\sigma}d\sigma\right).
\label{Wilson loop in CS}
\end{align}
The expectation value of the Wilson loop, $\langle W_R({\cal K}) \rangle_{CS}$, 
% \begin{align}
% \langle W_R({\cal K}) \rangle_{CS}=\frac{1}{{\cal Z}_{CS}}\int {\cal D}A\; W_R({\cal K})\; e^{iS_{CS}}
% \end{align}
defines a topological invariant
of ${\cal K}$.
%also depending on a choice of framing. 
In this letter, we mainly consider the Wilson loop for an unknot in the fundamental representation, 
denoted by $W_{\square}(\mbox{unknot})$, where $\square$ stands for the fundamental representation.
%While the canonical framing seems standard in the literature, 
It turns out that our formalism adopts
a non-canonical framing corresponding to the one labeled by $m=n=1$ in \cite{Blau:2006gh}.
% that also appears 
% in a direct evaluation of the partition function (\ref{partition function of CS}) 
% and corresponds to $m=n=1$ in the notation of \cite{Blau:2006gh}.
%and mainly consider the Wilson loop for the unknot in the fundamental representation, which we simply denote
%by $W(\mbox{unknot})$.
% For a link ${\cal L}$ with components ${\cal K}_{\alpha}\;\;(\alpha=1,\cdots,L)$, 
% the correlation function
% $\langle W_{R_1}^{{\cal K}_1}(A)\cdots W_{R_L}^{{\cal K}_L}(A) \rangle$
% is a topological invariant of ${\cal L}$.
% In this paper, we are interested in the case in which $L=1$ and ${\cal K}$ is a unknot.

In our framing, 
the above quantities are given by \cite{Witten:1988hf,Blau:2006gh,Marino:2004uf}
\begin{align}
&{\cal Z}_{CS}
=e^{-\frac{1}{12}g_sk(N^2-1)}\left(\frac{g_s}{2\pi i}\right)^{N/2}
\prod_{\alpha>0}2\sinh\frac{g_s\alpha\cdot\rho}{2} 
\label{partition function of CS 2} \\
&\;\;\;\;\;\;\;\;=\int\prod_{i=1}^N\frac{d\beta_i}{2\pi}
\prod_{i<j}\sinh^2\frac{\beta_i-\beta_j}{2}
e^{-\frac{1}{2g_s}\sum_i\beta_i^2}.
\label{partition function of CS 3} \\
&\langle W_{\square}(\mbox{unknot}) \rangle_{CS}
=e^{\frac{g_s}{2}\left(N-\frac{1}{N}\right)}\frac{\sinh\frac{g_sN}{2}}{\sinh\frac{g_s}{2}} 
=\sum_{i=1}^N \left\langle e^{\beta_i}\right\rangle_{CSM},
\label{vev of Wilson loop}
\end{align}
where $\rho$ is the Weyl vector of $SU(N)$, $\alpha>0$ are positive roots,
and we have introduced 
%\begin{align}
$g_s=2\pi i/(k+N)$,
%\label{g_s}
%\end{align}
which is identified with the string coupling in topological strings.
We have ignored a $g_s$-independent factor in rewriting (\ref{partition function of CS 2})
to (\ref{partition function of CS 3}).
$\langle \cdots \rangle_{CSM}$ denotes
the expectation value with the weight given by the integrand in
(\ref{partition function of CS 3}).

\paragraph*{Dimensional reduction.---}
In order to dimensionally reduce CS theory on $S^3$ \cite{Ishii:2007sy,Ishiki:2008vf},
we regard $S^3$ as the $SU(2)$ group manifold, which has the isometry 
%$SO(4)=SU(2)_L\times SU(2)_R$. 
$SO(4)=SU(2) \times SU(2)$ corresponding to the left and right translations.
We set the radius of $S^3$ to $2/\mu$, and
define the right-invariant 1-forms $E^i\;(i=1,2,3)$, 
which satisfy
the Maurer-Cartan equation
%\begin{align}
$dE^i-\frac{\mu}{2}\epsilon_{ijk}E^j\wedge E^k=0$.
%\label{Maurer-Cartan}
%\end{align}
We define the Killing vector ${\cal{L}}_i$ that is dual to $E^i$ and 
%corresponds to $SU(2)_L$.
%generators of the left translation, $SU(2)$. 
generates $SU(2)$ of the left translation.
%one of the two $SU(2)$'s.
%by ${\cal{L}}_i=-\frac{i}{\mu}E^M_i\partial_M$ 
%with $E^M_i$ being the inverse of $E^i_M$, where $M$ label three angular 
%variables.
We can also regard $S^3$ as an $S^1(U(1))$ bundle over $S^2=SU(2)/U(1)$.
The Kaluza-Klein (KK) momenta $q$ in the fiber direction $S^1$ 
take integers and half-integers. 
When ${\cal L}_i$ act on a KK mode with the momentum $q$, it takes 
the form of the angular momentum operators $L^{(q)}_i$
%in $R^3$ 
in the presence of a monopole 
with magnetic charge $q$ \cite{L_i}.
%at origin 
Namely, the KK momenta 
%along 
in the fiber direction are identified with the monopole charges on $S^2$.
%We simply denote $L^{(0)}_i$ by $Li$. 
%For the explicit forms of ${\cal L}_i$ and $L^{(q)}_i$, see \cite{b}.

Expanding the gauge field in (\ref{Chern-Simons theory on S^3}) as $A=i\mu X_iE^i$,
we rewrite RHS of (\ref{Chern-Simons theory on S^3}) as
\begin{align}
\!\!\!\!\!
-\frac{2k}{\pi}\!\int \!\! d\Omega_3
\mbox{Tr}\left(i\epsilon^{ijk}\left(X_i{\cal L}_jX_k 
+\frac{2}{3}X_iX_jX_k\right)+X_i^2\right),
\label{CS on S^3}
\end{align}
where $d\Omega_3$ is the volume element of unit 3-sphere, and we have used the Maurer-Cartan equation.
By only keeping the $q=0$ modes
in (\ref{CS on S^3}), namely replacing ${\cal L}_i$ and $d\Omega_3$ with
$L^{(0)}_i$ and $d\Omega_2$, respectively, we can dimensionally reduce the theory onto $S^2$.
The resultant theory is 
Yang-Mills theory (YM) on $S^2$.
To see this, we decompose $X_i$ into the radial component $\chi$ and the
components tangential to $S^2$, $a_{\theta}$ and $a_{\varphi}$,
%and as 
%$\vec{X}=\chi\vec{e}_r+a_{\theta}\vec{e}_{\varphi}-a_{\varphi}\vec{e}_{\theta}/\sin\theta$,
%\cite{Kitazawa:2002xj,Ishii:2008tm}, 
%where
%$r$, $\theta$ and $\varphi$ are the polar coordinates for $R^3$ 
% and $\vec{e}_{a}\;\;(a=r,\theta,\varphi)$ are
% the unit vectors in the corresponding directions,
and rewrite the theory on $S^2$ as
%\begin{align}
$\int d\Omega_2
\mbox{Tr}\left(\chi\epsilon^{\mu\nu} f_{\mu\nu}-\chi^2\right)$,
% $\frac{2\pi i}{g_{YM}^2A}\int d\Omega_2
% \mbox{Tr}\left(\chi\epsilon^{\mu\nu} f_{\mu\nu}-\chi^2\right)$,
%\label{BF with mass term on S^2 2}
%\end{align}
where 
%$g_{YM}^2=i\mu^2/(2k)$, $A=4\pi/\mu^2$ is the area of the $S^2$ and
$f_{\mu\nu}\;(\mu,\nu=\theta,\varphi)$ is the field strength for $a_{\mu}$. 
%The first term is the BF term and the second term is a mass term.
Integrating $\chi$ out indeed yields YM on $S^2$.
Finally, dropping all the derivatives in (\ref{CS on S^3}), namely
dimensionally reducing the theory onto a point gives rise to
a three-matrix model:
\begin{align}
S_{m}=-\frac{1}{g_m^2}\mbox{Tr}\left(X_i^2+\frac{i}{3}\epsilon^{ijk}X_i[X_j,X_k]\right).
\label{N=1^* matrix model}
\end{align}
%where $g_m^2=1/(4\pi k)$. 
%In the context of the Dijkgraaf-Vafa theory \cite{Dijkgraaf:2002fc}, 
%which takes the form of the superpotential for the so-called ${\cal N}=1^*$ theory. 
% We call the matrix model (\ref{N=1^* matrix model}) the ${\cal N}=1^*$ matrix model in this Letter.

\paragraph*{YM on $S^2$ from the matrix model.---}
The matrix model (\ref{N=1^* matrix model}) with the matrix size $M\times M$
possesses the following classical solutions,
\begin{align}
\hat{X}_i=\bigoplus_s L^{[j_s]}_i \otimes \idmat_{N_s},
\label{matrix background}
\end{align}
where $L^{[j]}_i$ are the spin $j$ representation of the $SU(2)$ generators, 
%obeying $[L^{[j]}_i,L^{[j]}_j]=i\epsilon_{ijk}L^{[j]}_k$, 
the relation $\sum_{s}(2j_s+1)N_s=M$ is 
satisfied, and $j_s\neq j_t$ for $s\neq t$. The label $s$ runs over some integers.

The $(s,t)$ block of the fluctuation around (\ref{matrix background}) is expanded in terms
of the fuzzy spherical harmonics \cite{Grosse:1995jt,Ishiki:2006yr,Ishii:2008tm}
which is a $(2j_s+1)\times(2j_t+1)$ matrix.
We put $2j_s+1=\Omega+n_s$ with $\Omega$ and $n_s$ being integers and take the limit in which 
\begin{align}
\Omega\rightarrow\infty, \;\;\; 
g_m^2/\Omega=-g_{YM}^2A/(8\pi^2)\;\;\mbox{fixed}
\label{limit}
\end{align}
with $g_{YM}^2=-\mu^2/(2k)$. 
$A=4\pi/\mu^2$ is the area of the $S^2$, and
$g_{YM}$ turns out to be the coupling constant of YM on $S^2$.
In this limit, the above fuzzy spherical harmonics coincides classically
with the monopole harmonics \cite{Wu:1976ge} of the monopole charge
$j_s-j_t$ under the identification $L_i^{[j_s]}\bullet-\bullet L_i^{[j_t]}\leftrightarrow L_i^{(j_s-j_t)}\bullet$ and
$\frac{1}{\Omega}\mbox{Tr}\leftrightarrow
\frac{1}{4\pi}\int d\Omega_2$. 
It was indeed shown in \cite{Ishii:2007sy} that
the theory around (\ref{matrix background}) is classically equivalent to the theory around
the following classical solution of $U(K)$ YM on $S^2$,
\begin{align}
\hat{X}_i=\mbox{diag}(\cdots,\underbrace{L^{(q_s)}_i-L^{(0)}_i
,\cdots,L^{(q_s)}_i-L^{(0)}_i}_{N_s}, \cdots)
\label{background of BF}
\end{align}
expressed in terms of $X_i$,  
% \begin{align}
% &\hat{\chi}=-\mbox{diag}(\cdots,\underbrace{q_s,\cdots,q_s}_{N_s}, \cdots),\nonumber\\
% &\hat{a}_{\mu}=\mbox{(single monopole config. with the unit charge)}\times\hat{\chi},
% \label{monopole solution}
% \end{align}
% given in terms of $a_{\mu}$ and $\chi$, 
where $q_s=n_s/2$, and the relation $\sum_sN_s=K$ is satisfied. 
% Indeed, the $(s,t)$ block of the fluctuation around (\ref{background of BF})
% just behaves as the one around (\ref{matrix background}) in the limit (\ref{limit}).

The above relationship is extended to the quantum level \cite{Ishiki:2008vf}.
The partition function of the matrix model (\ref{N=1^* matrix model}) is
decomposed into sectors classified by the representation of $SU(2)$ as in (\ref{matrix background}), and 
in the limit (\ref{limit}) the sectors consisting of $K$ blocks reduces to
% The partition function of the matrix model (\ref{N=1^* matrix model}) is
% decomposed into sectors classified by the representation of $SU(2)$ as in (\ref{matrix background}), and
% the sectors consisting of $K$ blocks reduces in the limit (\ref{limit}) up to an irrelevant constant to
\begin{align}
&\sum_{\{n_s,N_s\}}\int\prod_{s}\prod_{i=1}^{N_s}dy_{si} 
\prod_{s\leq t}\prod_{i=1}^{N_s}\prod_{j=1}^{N_t}
\left\{(y_{si}-y_{tj})^2-(n_s-n_t)^2\right\} \nonumber\\
&\;\;\;\;\;\;\;\;\;\;\;\;\times e^{-\frac{2\pi^2}{g_{YM}^2A}\sum_{s}\sum_{i=1}^{N_s}(y_{si}^2+n_s^2)}
\label{partition function of YM on S^2}
\end{align}
up to an irrelevant constant
with an analytic continuation $g_{YM}^2\rightarrow -ig_{YM}^2$. $y_{si}$ come from the eigenvalues of $X_3$.
This exactly agrees with the
partition function of $U(K)$ 2d YM on $S^2$ \cite{2d YM,note on U(1) part},
and the $\{n_s,N_s\}$ sector in the summation in (\ref{partition function of YM on S^2})
is the contribution around the classical background (\ref{background of BF}).

\paragraph*{CS theory on $S^3$ from the matrix model.---}
CS theory on $S^3$ is obtained by summing all the KK modes in YM on $S^2$, where
the KK momenta are identified with the monopole charges on $S^2$.
In the limit (\ref{limit}),
the $(s,t)$ block of the fluctuation around (\ref{matrix background}) of the matrix model
(\ref{N=1^* matrix model})  behaves as
the one around (\ref{background of BF}) of YM on $S^2$
 which feels the monopole charge $j_s-j_t$.
Therefore we are led to the following statement:
if one chooses in (\ref{matrix background}) 
\begin{align}
&-\Lambda/2 \leq s \leq \Lambda/2, \;\;
2j_s+1=\Omega+s, \;\; N_s=N,
\label{background}
\end{align}
with $\Lambda$ being a positive even integer
and takes the limit in which
\begin{align}
&\Omega\rightarrow\infty, \;\;\; \Lambda\rightarrow\infty, \;\;\; \Omega-\frac{\Lambda}{2}\rightarrow\infty, \;\;\;
N\rightarrow\infty, \nonumber\\
&g_m^2N/\Omega=N/(4\pi (k+N))\;\;\mbox{fixed},
\label{S^3 limit}
\end{align}
the theory around (\ref{background}) of the matrix model (\ref{N=1^* matrix model}) is equivalent 
to $U(N)$ CS theory on $S^3$ in the planar limit.  The naive relation between the coupling constants
is $g_m^2N/\Omega=N/(4\pi k)$, but we will see below that this naive one is renormalized so that
the last equation in (\ref{S^3 limit}) is valid.
The particular background (\ref{background}) and the limit (\ref{S^3 limit}) 
is the same as the ones adopted in \cite{Ishii:2008ib} in realizing
${\cal N}=4$ SYM on $R\times S^3$ in PWMM. 
This extraction of the theory around 
%the background 
(\ref{background}) from the whole theory of the matrix model
should be allowed in the $N\rightarrow\infty$ limit
as in the case of PWMM.
The above equivalence was classically shown in \cite{Ishii:2007sy} by further
imposing the orbifolding condition, which needs infinitely large matrix size 
and is not allowed quantum mechanically.
Here we do not need to impose it because of the $N\rightarrow\infty$ limit.
$\Omega$ and $\Lambda$ are viewed as the ultraviolet cutoffs for the angular momenta,
which remarkably respect the gauge symmetry.

More precisely, the equivalence states that 
\begin{align}
{\cal F}_m/(N^2(\Lambda+1))={\cal F}_{CS}/N^2
\label{free energy equivalence}
\end{align}
in the limit (\ref{S^3 limit}),
where ${\cal F}_m$ is the free energy of the theory around the background (\ref{background}) of the
matrix model (\ref{N=1^* matrix model}), and ${\cal F}_{CS}=\ln {\cal Z}_{CS}$.
Furthermore, given a knot ${\cal K}$ in $S^3$, 
we can parametrize it by $z^M(\sigma) \;\;(\sigma\in[0,1])$ and
introduce a Wilson loop in the matrix model,
\begin{align}
\hat{W}({\cal K})=\frac{1}{M}{\rm Tr} \left[
P\exp \left( i\mu \int^1_0 X_iE^i_M(z(\sigma))
\frac{dz^M(\sigma)}{d\sigma}d\sigma
\right)
\right]. 
\nonumber
%\label{Wilson loop in matrix model}
\end{align}
It was shown classically in \cite{Ishii:2007sy} that
the above operator around 
%the background 
(\ref{background}) in the limit (\ref{S^3 limit})
%is equivalent to the corresponding Wilson loop (\ref{Wilson loop in CS}), $W_{\square}({\cal K})$.
corresponds to $W_{\square}({\cal K})$ in (\ref{Wilson loop in CS}).
%in the fundamental representation.
%in CS theory on $S^3$ . 
We naturally expect this correspondence to hold also at the quantum level:
\begin{align}
\langle\hat{W}({\cal K})\rangle_m=\langle W_{\square}({\cal K})\rangle_{CS}/N,
\label{Wilson loop relation}
\end{align}
where $\langle\cdots\rangle_m$
denotes the expectation value
around the background (\ref{background}) of the matrix model in the limit (\ref{S^3 limit}).
Here we concentrate on an unknot whose path is given by a great circle on $S^3$.
In this case, the relation (\ref{Wilson loop relation}) reduces to
%We parametrize the path  by
%$z^M=(\theta, \varphi, \psi)=\left(\frac{\pi}{2},0, 4\pi\sigma\right)$ 
%using the notation of Ref. \cite{Ishii:2007sy}. 
%Substituting this parametrization into (\ref{Wilson loop relation})
%gives rise to
\begin{align}
\left\langle{\rm Tr} \left(e^{ 4\pi i X_3 }\right)\right\rangle_m/M
=\langle W_{\square}(\mbox{unknot})\rangle_{CS}/N.
\label{Wilson loop equivalence}
\end{align}

\paragraph*{Proof of the equivalence.---}
To prove the equivalence, we show the equalities (\ref{free energy equivalence}) and 
(\ref{Wilson loop equivalence}) explicitly. For that purpose, we may use the statistical model whose partition
function is defined by
(\ref{partition function of YM on S^2}), by taking the $\Omega\rightarrow\infty$ limit first.
We extract the contribution of $-\Lambda/2\leq s \leq \Lambda/2$, $n_s=s$,
$N_s=N$ from the sum in (\ref{partition function of YM on S^2}).
%By considering (\ref{limit})
%and (\ref{S^3 limit}) and making an analytical continuation
%$y_{si}\rightarrow -i\sqrt{\lambda}\phi_{si}/(2\pi)$, where $\lambda=g_sN$ is the 't Hooft coupling,
Then, we see that the model is interpreted as
a multi-matrix model of $N\times N$ hermitian matrices with double trace interactions 
in which $y_{si}$ is a constant times the $i$-th eigenvalue of the
$s$-th matrix $\phi_s$:
\begin{align}
{\cal Z}=\int \prod_sd\phi_s \exp\left(-\frac{N}{2}\sum_s\mbox{tr}\phi_s^2-\tilde{V}\right),
\label{multi-matrix model}
\end{align}
where
\begin{align}
\tilde{V}
&=\sum_{s\neq t}\sum_{k=1}^{\infty}\sum_{m=0}^{2k}\frac{1}{2k(s-t)^2}\left(\frac{-\lambda}{4\pi^2}\right)^k 
\left(\begin{array}{c}
2k \\
m
\end{array}\right) \nonumber\\
&\qquad\qquad\qquad\;\;\;\;\;\times\mbox{tr}\phi_s^m\mbox{tr}(-\phi_t)^{2k-m}
\nonumber
\end{align}
with $\lambda=g_sN=2\pi i N/(k+N)$. (\ref{S^3 limit}) tells us to 
take the limit in which $\Lambda\rightarrow\infty,\;N\rightarrow\infty$ 
with $\lambda$ fixed.
Recalling the computation in obtaining (\ref{partition function of YM on S^2})  
from the matrix model (\ref{N=1^* matrix model})
in \cite{Ishiki:2008vf}, we find that 
LHS of (\ref{Wilson loop equivalence}) is rewritten as 
$\langle \sum_s\mbox{tr}e^{\sqrt{\lambda}\phi_s}/(N(\Lambda+1)) \rangle$, where $\langle\cdots\rangle$ denotes
the expectation value in (\ref{multi-matrix model}).
In the same way, by using the formula $\prod_{n=1}^{\infty}(1+x^2/n^2)=\sinh(\pi x)/(\pi x)$,
%and $\zeta(s)=\sum_{n=1}^{\infty}1/n^s$,
we also see that the statistical model defined by (\ref{partition function of CS 3}) 
is interpreted as a one-matrix model of an $N\times N$ hermitian matrix $\phi$
with double-trace interaction terms \cite{Aganagic:2002wv}:
\begin{align}
{\cal Z}_{M}=\int d\phi \exp\left(-\frac{N}{2}\mbox{tr}\phi^2-V \right),
\label{one-matrix model}
\end{align}
where
\begin{align}
V=\sum_{k=1}^{\infty}\sum_{m=0}^{2k}\frac{\zeta(2k)}{k}\left(\frac{-\lambda}{4\pi^2}\right)^k
\left(\begin{array}{c}
2k \\
m
\end{array}\right)
\mbox{tr}\phi^m\mbox{tr}(-\phi)^{2k-m}.
\nonumber
\end{align}
% \begin{align}
% V=\sum_{k=1}^{\infty}\sum_{m=0}^{2k}\frac{\zeta(2k)}{k}\left(\frac{-\lambda}{4\pi^2}\right)^k
%             {}_{2k}C_m\mbox{tr}\phi^m\mbox{tr}(-\phi)^{2k-m}.
% \nonumber
% \end{align}
Noting (\ref{vev of Wilson loop}), we finally see that
(\ref{free energy equivalence}) and (\ref{Wilson loop equivalence}) reduce to
\begin{align}
&{\cal F}/(N^2(\Lambda+1))
={\cal F}_{M}/N^2, 
\label{free energy equivalence 2} \\
&\frac{1}{N(\Lambda+1)}\sum_s\left\langle\mbox{tr} e^{\sqrt{\lambda}\phi_s}\right\rangle
=\frac{1}{N}\langle\mbox{tr}e^{\sqrt{\lambda}\phi}\rangle_{M},
\label{Wilson loop equivalence 2}
\end{align}
respectively, where ${\cal F}=\ln {\cal Z}$, ${\cal F}_{M}=\ln {\cal Z}_{M}$, and
$\langle\cdots\rangle_{M}$ denotes the expectation value in (\ref{one-matrix model}).

To show (\ref{free energy equivalence 2}) and (\ref{Wilson loop equivalence 2}),
we perform the perturbative expansion in $\lambda$ in both the matrix models using the standard double line
notation. 
We represent the vertices by connecting the double traces in terms of dashed lines as in Fig. 1.
There is one-to-one correspondence between the diagrams of the two matrix models.
As in the standard perturbation theory, the free energy and the expectation value of the Wilson loop
are computed by summing all the connected diagrams,
where the diagrams that have pieces connected in terms of the dashed lines are regarded as connected.
It is easy to see that the leading contribution in the $1/N$ expansion
is given by the diagrams that are planar in the ordinary
sense and in addition 
%`tree' with respect to
%the dashed lines. 
divided into two parts by cutting any dashed line.
The diagrams that do not satisfy these two conditions
%in which the dashed lines form any loop 
give the subleading contribution in the $1/N$ expansion.
Namely, only such `tree' planar diagrams survive 
%in the planar limit.
in the $N\rightarrow\infty$ limit.
%of CS theory on $S^3$.
Fig. 1 shows examples of the diagrams for the free energy: the left one is `tree' planar and
gives the leading contribution,
while the right one gives the subleading contribution.
%an example of the `tree' planar diagrams 
%giving the leading contribution 
%in $1/N$ 
%(left) and an example of the diagrams giving the subleading contribution 
%in $1/N$ 
%(right)
%in 
%the calculation 
%perturbative expansion 
%of 
%the free energy.
%In the case of 
For the left diagram in Fig. 1, we can prove an equality 
% \begin{align}
% \frac{1}{2^2(\Lambda+1)}\sum_{s,t,u}\frac{1}{(s-t)^6(s-u)^4}=\zeta(6)\zeta(4)
% \end{align}
\begin{align}
&\frac{1}{2^4(\Lambda+1)}\sum_{s,t,u,v,w}\frac{1}{(s-t)^{8}(t-u)^{10}(u-v)^4(u-w)^4} \nonumber\\
&=\zeta(8)\zeta(10)\zeta(4)\zeta(4),
\nonumber
\end{align}
in the $\Lambda\rightarrow\infty$ limit, so that
we find that the diagram takes the same value in the two matrix models.
Such an equality does not hold 
%in the case of 
for the right diagram in Fig. 1, so that the diagram takes different values
in the two matrix models.
For a generic `tree' planar diagram shown in Fig. 2 possessing $n$ dashed lines, 
%and gives leading order contribution in $1/N$,
we can prove the following equality in the $\Lambda\rightarrow\infty$ limit \cite{full paper}:
\begin{align}
\frac{1}{2^n(\Lambda+1)}\sum_{\{s_a\}}\frac{1}{(s_1-s_2)^{2k_1}\cdots(s_{n}-s_{n+1})^{2k_n}} 
\!=\!\prod_{a=1}^n\zeta(2k_a),
\nonumber
\end{align}
which implies that the diagram takes the same value in the two matrix models.
Thus we have shown that (\ref{free energy equivalence 2}) and (\ref{Wilson loop equivalence 2}) hold
to all order in $\lambda$ in the limit in which  $\Lambda\rightarrow\infty$ and $N\rightarrow\infty$. 
%(\ref{S^3 limit}).
%This expansion has a finite radius of convergence, so that 

% \begin{figure}[htb]
% \begin{center}
% \includegraphics[height=1cm]{diagram1.eps}
% \end{center}
% \caption{Vertex including $\mbox{tr}(\phi^6)\mbox{tr}(\phi^4)$ or $\mbox{tr}(\phi_s^6)\mbox{tr}(\phi_t^4)$}
% \label{vertex}
% \end{figure}

% \begin{figure}[tb]
% \begin{center}
% \hspace*{-2cm}
%  \begin{minipage}{0.1\hsize}
%   \begin{center}
% \includegraphics[height=1.4cm]{tree.eps}
% \end{center}
%  \end{minipage}
% \hspace*{4cm}
% \begin{minipage}{0.1\hsize}
%   \begin{center}
% \includegraphics[height=1.4cm]{nontree.eps}
% \end{center}
%  \end{minipage}
% \caption{A `tree' planar diagram for the free energy  (left); 
% A `non-tree' planar diagram for the free energy (right).}
% \end{center}
% \label{tree and nontree}
% \end{figure}

\begin{figure}[tb]
\begin{center}
\includegraphics[height=2cm]{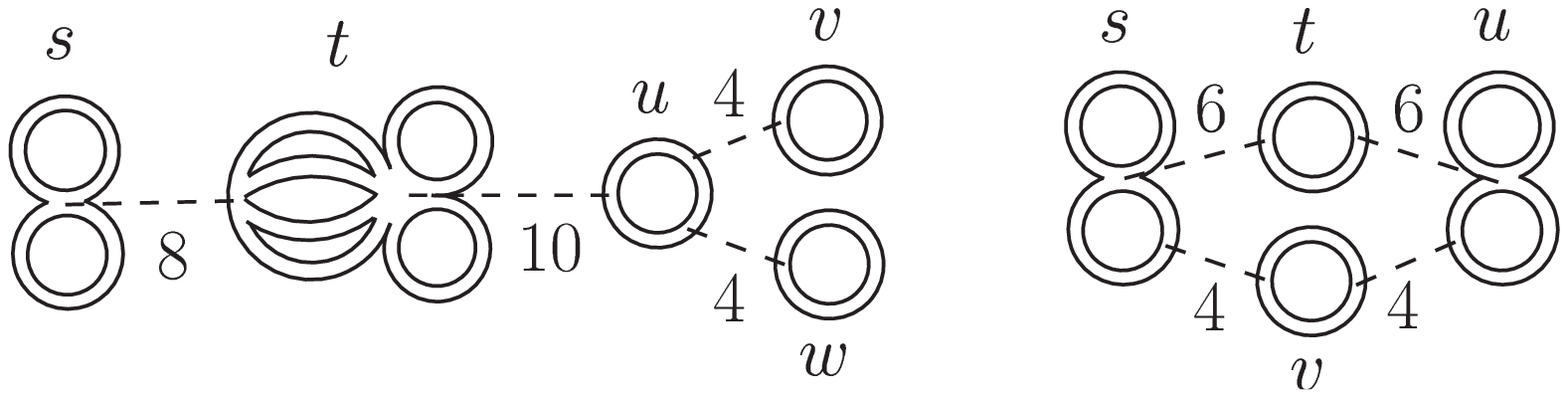}
\end{center}
\caption{Diagrams for the free energy: A `tree' planar diagram giving the leading contribution 
in the $1/N$ expansion (left) and a diagram giving the subleading contribution (right). Each number
represents the degree of the vertex.} 
\label{tree and non-tree diagram}
\end{figure}

\begin{figure}[tb]
\begin{center}
\includegraphics[height=2cm]{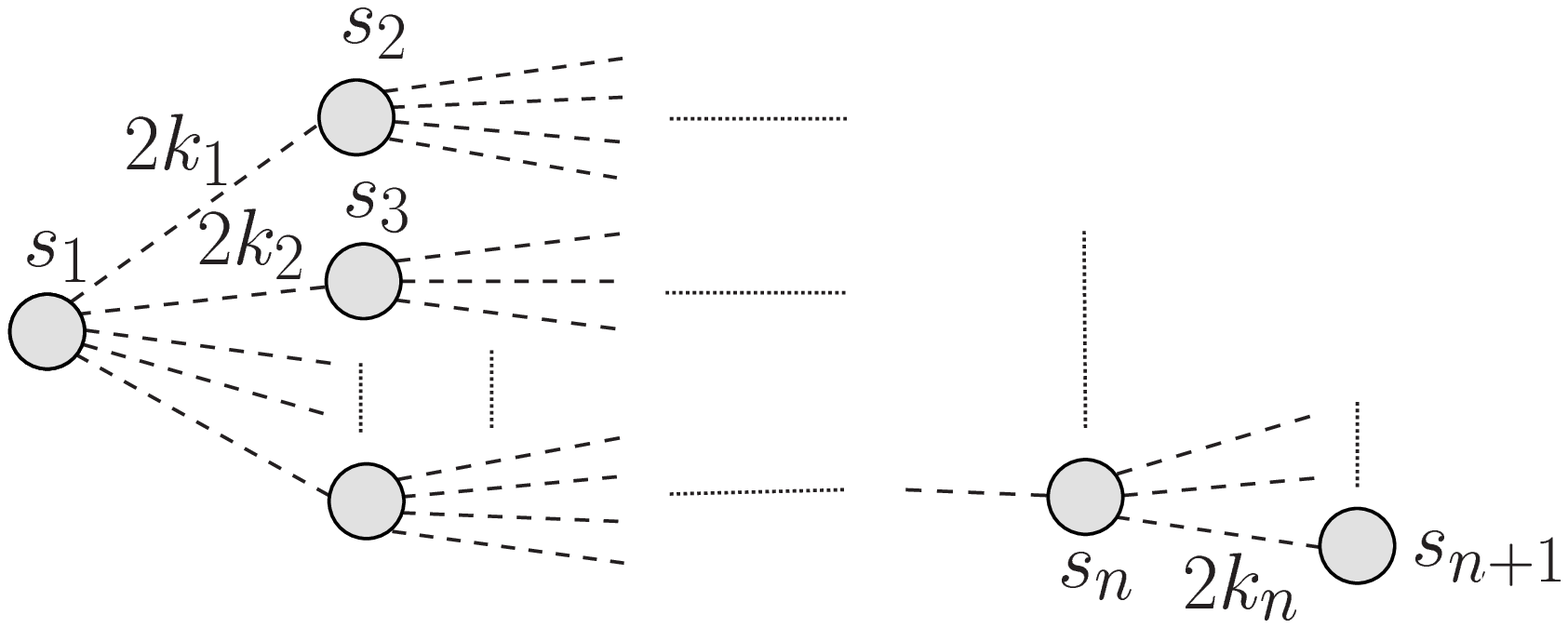}
\end{center}
\caption{A generic `tree' planar diagram for the free energy. Each blob represents a planar diagram.} 
\label{generic tree diagram}
\end{figure}
%%%%%%%%%%%%%%%%%%

\paragraph*{Summary.---}
In this letter, we showed that expanded around the background (\ref{background}) in the limit (\ref{S^3 limit}), the matrix model is equivalent to
planar $U(N)$ CS theory on $S^3$.
%which is viewed
%considered 
%as a new type of the large-$N$ reduction.
We can also show that expanded around different backgrounds, the matrix model reproduces planar CS theory
on $S^3/Z_q$,
%in appropriate limits, 
which we will describe in 
%another publication 
\cite{full paper}.
Our formalism does not only give a new regularization method of 
planar CS theory on $S^3$, but also serves as 
the first proved example of the new type of the large $N$ reduction.

\paragraph*{Acknowledgments.---}
We would like to thank K. Ohta for his collaboration in the early stage of this work
and for many discussions.
The work of S.\ S.\ 
is supported 
%in part 
by JSPS.
%is supported in part by the JSPS Research Fellowship for Young Scientists. 
The work of A.\ T.\ 
is supported 
%in part 
by Grant-in-Aid for Scientific Research (19540294) from JSPS.

%\end{references}

\end{document}